\documentclass[aps,pre,amsmath,groupedaddress,amssymb,showpacs,preprint,showkeywords]{revtex4}
\usepackage{dcolumn}
\usepackage[dvips]{graphicx}
\usepackage[utf8]{inputenc}

\begin{document}

\title{Newtonian Dynamics from the principle of Maximum Caliber}

\author{Diego González}
\email{dgonzalez@gnm.cl}

\author{Sergio Davis}
\email{sdavis@gnm.cl}

\author{Gonzalo Guti\'errez}
\email{gonzalo@macul.ciencias.uchile.cl}

\affiliation{Grupo de Nanomateriales, Departamento de F\'{\i}sica, Facultad de Ciencias,
Universidad de Chile, Casilla 653, Santiago, Chile}

\date{\today}

\begin{abstract}
The foundations of Statistical Mechanics can be recovered almost in their
entirety from the Principle of Maximum Entropy. In this work we show that its non-equilibrium 
generalization, the Principle of Maximum Caliber (Jaynes, 1980), when applied to
the unknown trajectory followed by a particle, leads to Newton's second law under two quite intuitive 
assumptions (the expected square displacement in one step and the spatial probability distribution of the
particle are known at all times). Our derivation explicitly highlights the role of mass 
as an emergent measure of the fluctuations in velocity (inertia) and the origin of potential energy 
as a manifestation of spatial correlations. According to our findings, the application of Newton's
equations is not limited to mechanical systems, and therefore could be used in
modelling ecological, financial and biological systems, among others.
\end{abstract}



\maketitle

\section{Introduction}

In 1957, E. T. Jaynes ~\cite{Jaynes1957} postulated that Statistical Mechanics 
has to be understood, not as a physical theory in the same footing as, say, classical 
mechanics or electromagnetism, but as an application of statistical inference on
a system with macroscopically large numbers of degrees of freedom. The question
was reversed from ``given the microscopic evolution of the system, what is the
probability distribution for the macroscopic quantities?'' to ``given a few known
macroscopic properties, what are the possible microstates compatible with said
knowledge?''. The answer, as initially proposed by Gibbs, was the probability
distribution with maximum entropy $S=-\sum_i P_i\ln P_i$ subjected to constraints 
reflecting the known macroscopic properties. Jaynes, after the work of Shannon
in information theory, realized that this procedure (maximization of $S$
constrained only by the known information) is not limited to Statistical Mechanics but 
a valid principle in any problem of statistical inference. Due to the uniqueness of 
Shannon's entropy in characterizing uncertainty it is the most unbiased procedure for the
construction of statistical models. Later, it has been axiomatically
derived~\cite{Shore1980, Skilling1988} from requirements of internal consistency.

The principle of Maximum Caliber~\cite{Jaynes1980} generalizes the idea of
Maximum Entropy to dynamical systems, including time explicitly. For this, we
now ask for the possible microscopical trajectories compatible with known
information. The result is the probability distribution of trajectories
$P[x(t)|H]$ which maximizes the Shannon entropy, now defined as the functional
integral

\begin{equation}
\mathcal{S} = -\int Dx(t) P[x(t)|H] \ln P[x(t)|H].
\end{equation}

Maximum Caliber has been applied recently to discrete dynamics~\cite{Stock2008} and earlier to derive the
Fokker-Planck equations~\cite{Haken1986} and the Markov process formalism~\cite{Ge2011}.

Every Maximum Caliber solution predicts that the most probable trajectory is the
one that extremizes some functional (analogous to an ``action'' in classical
mechanics), in the same way that every Maximum Entropy solution predicts that
the most probable microstate is the one that extremizes some function which is a
combination of all the constraints imposed. This leads to the question: without
introducing the Lagrangian of classical mechanics explicitly, could it
``emerge'' naturally from simpler constraints in a Maximum Caliber problem?

Newtonian dynamics has been previously derived from information-geometric
arguments~\cite{Caticha2007} leading to the idea of entropic dynamics. This idea
is based on the assumption of an irreducible uncertainty in the position of a
particle, implying an information metric for space from which Newton's second
law naturally emerges. Caticha's derivation is founded on the Maximum Entropy
principle, suitably modified to update a prior distribution under new constraints.

In this work we show that if we use the Maximum Caliber principle to find the unknown 
trajectory of a particle, there are two general conditions that lead to Newton's second 
law, namely that (a) the expected square displacement per step is known at all times, and 
(b) that the time-independent probability of finding the particle at any coordinate is also known. 
Knowledge of both (a) and (b) leads to Newton's second law in expectation over
trajectories, and what is perhaps more interesting, any dynamical system not following 
Newton's second law has to violate at least one of these assumptions.

\section{The Maximum Entropy and Maximum Caliber formalism}

Consider a system with $N$ degrees of freedom, whose states are denoted by
vectors $\vec{x}=(x_1,\ldots, x_N)$. Suppose the expectation values of $M$
functions $f_i(\vec{x})$ are known. Maximization of the Shannon entropy leads to 
the MaxEnt model

\begin{equation}
P(\vec{x}|H) = \frac{1}{Z(\vec{\lambda})}\exp\left(-\sum_{k=1}^M \lambda_k
f_k(\vec{x})\right),
\label{eq_maxent}
\end{equation}
where the value of the Lagrange multipliers $\lambda_k$ needed to impose the $M$ 
constraints can be determined from

\begin{equation}
-\frac{\partial}{\partial \lambda_k} \ln Z(\vec{\lambda}) = \big<f_k(\vec{x})\big>.
\label{eq_lagrangemul}
\end{equation}

This nonlinear equation is usually impractical to solve, as it needs the
partition function explicitly. It has been recently shown that~\cite{Davis2012} for the Lagrange multipliers the
equality

\begin{equation}
\Big<\nabla\cdot \vec{v}\Big> = \sum_{k=1}^M \lambda_k \Big<\vec{v}\cdot \nabla f_k\Big>
\label{eq_cvt}
\end{equation}
holds, with $\vec{v}$ an arbitrary differentiable vector field, and this
provides a linear system of equations for $\vec \lambda$.

Now, suppose $N$ is infinitely large, the state vector $\vec{x}$ becomes a
function of a continuous parameter, let us call it $t$, i.e., a parameterized 
trajectory. The probability distribution functional for the different possible
trajectories is (from Eq. \ref{eq_maxent} in the $N\rightarrow \infty$ limit), 

\begin{equation}
P[x(t)|H] = \frac{1}{Z[\lambda(t)]}e^{-\int dt \lambda(t) f[x(t); t]},
\end{equation}
where, similarly to Eq. \ref{eq_lagrangemul}, the Lagrange multiplier function can be obtained from

\begin{equation}
-\frac{\delta}{\delta \lambda(t)} \ln Z[\lambda(t)] = \big<f[x(t); t]\big>.
\end{equation}

If we discretize time, the trajectory $x(t)$ becomes a vector $\vec{x}=(x_0, \ldots, x_{n-1})$, and the Lagrange function $\lambda(t)$ becomes a vector
$\vec \lambda=(\lambda_0, \ldots, \lambda_{n-1})$. In fact, we recover Eqs.
\ref{eq_maxent} and \ref{eq_lagrangemul}. This means we can use Eq. \ref{eq_cvt} in a discretized Maximum Caliber problem.

\section{Derivation of Newton's second law}

Consider a single particle following an unknown trajectory $x(t)$ in one spatial
dimension. This can be easily generalized to many particles in arbitrary
dimensions, at the cost of overcomplicated notation. We can discretize this
trajectory in $n$ steps, such that $x(t)$ now becomes a vector $\vec{x}=(x_0,
\ldots, x_{n-1})$, and then impose the following constraints (expectations are
to be interpreted over all possible trajectories) 

\begin{eqnarray}
\Big<(x_i - x_{i-1})^2\Big> = (\Delta t)^2{d_i}^2 \\
\Big<\delta(x_i-X)\Big> = P(x_i=X|H),
\end{eqnarray}
for all values of $i$ and $X$. The first constraint recognizes the fact that
the expected square displacement in one (possible infinitesimal) step is known for all times, and is equal to an arbitrary function
${d_i}^2$ times the time step. We expressed it in this form so that $d_i$ can
remain finite when taking the limit $\Delta t \rightarrow 0$. The second
constraint imposes that the static, time-independent probability distribution for 
the coordinate $x$ is also known.

The probability distribution function for $\vec{x}$ is

\begin{equation}
P(\vec x|H) = \frac{1}{Z(\vec
\lambda)}\exp\Big(-\sum_{i=0}^{n-1}\frac{\lambda_i}{(\Delta t)^2}
(x_i-x_{i-1})^2+\sum_{i=0}^{n-1}\int dX \mu(X)\delta(x_i-X)\Big)
\end{equation}
which, after integrating the Dirac delta function, becomes
 
\begin{equation}
P(\vec x|H) = \frac{1}{Z(\vec
\lambda)}\exp\Big(-\sum_{i=0}^{n-1}\frac{\lambda_i}{(\Delta t)^2} (x_i-x_{i-1})^2+\sum_{i=0}^{n-1}\mu(x_i)\Big).
\label{eq_prob}
\end{equation}

This is the probability of the particle taking a well-defined discretized
trajectory $\vec x$, and is precisely the solution of a Maximum Entropy problem with
$n$ degrees of freedom and $n$ Lagrange multipliers $\lambda_i$ (plus the
function $\mu$), therefore Eq. \ref{eq_cvt} holds as 

\begin{equation}
\Big<\nabla \cdot \vec{v}(\vec{x})\Big> =
\sum_{i=0}^{n-1}\frac{\lambda_i}{(\Delta t)^2}\Big<\vec{v}(\vec{x})\cdot \nabla (x_i-x_{i-1})^2\Big> +
\sum_{i=0}^{n-1}\Big<\vec{v}(\vec{x})\cdot \nabla \mu(x_i)\Big>,
\end{equation}
with $\vec{v}$ an arbitrary vector field, of our choosing. If we choose
$\vec{v}$ such that it has a single component $k$, i.e.
$v_i=\delta_{i,k}\omega(\vec{x})$ with $\omega$ an arbitrary scalar field, we
obtain

\begin{eqnarray}
\Big<\frac{\partial \omega}{\partial x_k}\Big> =
\sum_{i=0}^{n-1}\frac{\lambda_i}{(\Delta t)^2}\Big<\omega(\vec{x})\cdot
2(x_i-x_{i-1})(\delta_{i,k}-\delta_{i-1,k})\Big> +
\sum_{i=0}^{n-1}\Big<\omega(\vec{x})\cdot \mu'(x_i)\delta_{i,k}\Big> \\ \nonumber
=\frac{1}{(\Delta t)^2}\Big<2\omega(\vec{x})\Big[\lambda_k(x_k-x_{k-1})-\lambda_{k+1}(x_{k+1}-x_k)\Big]\Big> +
\Big<\omega(\vec{x})\mu'(x_k)\Big>.
\label{eq_newton_first}
\end{eqnarray}

But recalling that the discrete forward derivative is

\begin{equation}
\dot{a}_i \approx \frac{a_{i+1}-a_i}{\Delta t},
\end{equation}
we can write Eq. \ref{eq_newton_first} as 

\begin{equation}
\Big<\frac{\partial \omega}{\partial x_k}\Big> = -\Big<\omega\Big(\dot{p}_k + \mu'(x_k)\Big)\Big>,
\label{eq_newton_omega}
\end{equation}
where 
\begin{equation}
p_k=2\lambda_k\dot{x}_k=m_k\dot{x}_k.
\end{equation}

Considering $\omega=1$ and defining $\Phi(x)=-\mu(x)$ we finally obtain

\begin{equation}
\Big<\dot{p}_k\Big> = -\Big<\Phi'(x_k)\Big>.
\label{eq_newton}
\end{equation}
which is a discrete version of Newton's second law with momentum $p(t)=m(t)\dot{x}(t)$ and potential energy $\Phi(x)$.

From this we note that a time-dependent mass $m(t)$ and a potential energy have
emerged from the Lagrange multipliers associated with the constraints on
the expected square of the step and the probability distribution of the
coordinate, respectively. Thus we can say the following: whenever the
information about the expected square of the step is important, the particle
acquires mass, and whenever the information about which regions are more
probable in space becomes important, the particle is subjected to a potential energy.

The most probable trajectory for the particle follows a minimum action
principle. Indeed, if we replace our definitions of $m_k$ and $\Phi(x_k)$ in Eq.
\ref{eq_prob}, we recover in the exponential the classical action

\begin{equation}
P(\vec{x}|H) =
\frac{1}{Z}\exp\Big(-\sum_{i=0}^{n-1}\Big[\frac{1}{2}m_i{\dot{x}_i}^2 -
\Phi(x_i)\Big]\Big)
\end{equation}
which in the continuum limit becomes
\begin{equation}
P[x(t)|H] = \frac{1}{Z}\exp\Big(-\int dt
\mathcal{L}(t)\Big).
\end{equation}

This tells us that the most probable trajectory is the one that extremizes the
classical action with Lagrangian

\begin{equation}
\mathcal{L}(t) = \frac{p(t)^2}{2m(t)}-\Phi(x(t))
\end{equation}

and associated Hamiltonian 

\begin{equation}
\mathcal{H}=\frac{p(t)^2}{2m(t)}+\Phi(x(t)).
\label{eq_ham}
\end{equation}

Therefore the most probable trajectory is governed by the canonical formalism of
Classical Mechanics. In appendix A we explore the validity of some aspects of the canonical
formalism, namely the Poisson bracket, for the expectation over trajectories.

\section{Concluding remarks}

We have found that two simple constraints are sufficient to recover Newton's
second law in expectation for the probable trajectories of a particle. The first
constraint, on the step size as a function of time, leads to the existence of an
inertial mass $m(t)$ proportional to the Lagrange multiplier $\lambda(t)$. To
understand the meaning of this, remember that for any variational problem solved
using Lagrange multipliers, the larger the value of the multiplier, the more
restrictive (and therefore more relevant) the constraint. An irrelevant
constraint has always a vanishing multiplier. As Jaynes~\cite{Jaynes1982} (p.
945) clearly states, ``The Lagrange multipliers $\lambda_k$ in the MAXENT
formalism have therefore a deep meaning: $\lambda_k$ is the 'potential' of the
datum $R'_k$, that measures how important a constraint it represents.''

Now we motivate the following principle: constraints related to conserved 
quantities are always more relevant. For instance, this explains the fact that
the canonical ensemble in equilibrium statistical mechanics is correctly derived
just from a single constraint, the energy or expectation of the Hamiltonian,
which is an integral of motion. Another illustration is the following: suppose
we are trying to recover the trajectory of a particle from information about the
distance to a particular point. If this distance is a constant, this is enough
to isolate a unique trajectory, the circle. If we only know that the distance
varies between $r_1$ and $r_2$, the number of compatible trajectories will increase 
with $\Delta r=r_2-r_1$, thus the strength of the constraint will
correspondingly decrease with increasing $\Delta r$.

Given the earlier discussion, the closer $d_i^2$ is to be a conserved quantity,
the more relevant the first constraint is. In this case, $\lambda(t)$ is large and 
therefore, $m(t)$ is also large. Conversely, if the value of $m$ is small, this
means $\lambda(t)$ is small and therefore $d_i^2$ has larger fluctuations. In
the continuous limit it is the instantaneous speed that fluctuates (there is a non-zero 
acceleration). This embodies the idea of inertia, and is reminiscent of the
ideas of Smolin~\cite{Smolin1986} and of Nelson~\cite{Nelson1966} about inertia
being inversely proportional to the size of quantum fluctuations.

\section{Acknowledgements}

DG gratefully acknowledges the access to resources provided by Grupo de Nano
Materiales (Departamento de F\'{\i}sica, Facultad de Ciencias, Universidad de Chile).

\appendix

\section{Canonical coordinates and Poisson brackets}

An interesting question is how much of the formalism of classical mechanics we
can recover from Eq. \ref{eq_newton}. The fact that most of the structure of classical
mechanics is contained in the definition and properties of the Poisson bracket,
motivates us to search for an operation analogous to this bracket under the
Maximum Caliber formalism.

For arbitrary functions $f(x, p)$ and $g(x, p)$ the Poisson bracket is defined as 

\begin{equation}
\{f, g\} = \frac{\partial f}{\partial x}\frac{\partial g}{\partial p} -
\frac{\partial f}{\partial p}\frac{\partial g}{\partial x},
\end{equation}
and it is such that 

\begin{equation}
\frac{df}{dt}-\frac{\partial f}{\partial t} = \{f, \mathcal{H}\}
\end{equation}
holds. Let us compute the expectation of the left hand side,

\begin{equation}
\Big<\frac{df}{dt}\Big> - \Big<\frac{\partial f}{\partial t}\Big>  = \Big<\frac{\partial
f}{\partial x_k}\dot{x}_k+\frac{\partial f}{\partial p_k}\dot{p}_k\Big>,
\end{equation}
which using Eq. \ref{eq_newton_omega} with $\omega=\partial f/\partial p_k$ can
be written as 

\begin{equation}
\Big<\frac{df}{dt}\Big> - \Big<\frac{\partial f}{\partial t}\Big> = \Big<\frac{\partial
f}{\partial x_k}\dot{x}_k - \frac{\partial}{\partial x_k}\Big(\frac{\partial
f}{\partial p_k}\Big) - \frac{\partial f}{\partial p_k}\Phi'(x_k)\Big>
\end{equation}

Now using our classical Hamiltonian (Eq. \ref{eq_ham}) we recognize its derivatives 
\begin{eqnarray}
\dot{x}_k = \frac{\partial \mathcal{H}}{\partial p_k} \\
\Phi'(x_k) = \frac{\partial \mathcal{H}}{\partial x_k}
\end{eqnarray}
and, upon replacing, we have 

\begin{equation}
\Big<\frac{df}{dt}\Big> - \Big<\frac{\partial f}{\partial t}\Big> = \Big<\frac{\partial
f}{\partial x_k}\frac{\partial \mathcal{H}}{\partial p_k} - \frac{\partial}{\partial x_k}\Big(\frac{\partial
f}{\partial p_k}\Big) - \frac{\partial f}{\partial p_k}\frac{\partial
\mathcal{H}}{\partial x_k}\Big>
\end{equation}
leading finally to

\begin{equation}
\Big<\frac{df}{dt}\Big> - \Big<\frac{\partial f}{\partial t}\Big> = \Big<\{f, \mathcal{H}\}\Big> -
\Big<\frac{\partial}{\partial x_k}\Big(\frac{\partial
f}{\partial p_k}\Big)\Big>.
\end{equation}

So, in expectation we find a Poisson bracket analog with an additional term. For the particular case $f=\mathcal{H}$, we
obtain

\begin{equation}
\Big<\frac{d\mathcal{H}}{dt}\Big> = -\Big<\frac{\partial}{\partial
x_k}\Big(\frac{\partial \mathcal{H}}{\partial p_k}\Big)\Big>,
\end{equation}
which reduces to

\begin{equation}
\Big<\frac{d\mathcal{H}}{dt}\Big> = -\Big<\frac{\partial \dot{x}_k}{\partial
x_k}\Big> = 0,
\end{equation}
using the centered difference~\cite{OneSided},

\begin{equation}
\dot{a}_i \approx \frac{a_{i+1}-a_{i-1}}{2\Delta t}.
\end{equation}

Therefore we have shown that, for a Hamiltonian with the form given in Eq.
\ref{eq_ham}, the energy is conserved in expectation. 

\bibliography{newton}

\begin{thebibliography}{13}
\bibitem[{One()}]{OneSided}
\bibinfo{note}{Note that using one-sided differences is not appropriate here,
  as one obtains $\pm 1/\Delta t$ depending on forward or backward.}
\end{thebibliography}
\bibliographystyle{apsrev}

\end{document}